\documentclass[useAMS,usenatbib]{mn2e}
\usepackage{graphicx}
\def\lta{\lower2pt\hbox{$\buildrel {\scriptstyle <}
   \over {\scriptstyle\sim}$}}
\def\gta{\lower2pt\hbox{$\buildrel {\scriptstyle >}
   \over {\scriptstyle\sim}$}}
\def\apj{ApJ}
\def\nat{Nature}
\def\apjl{ApJ}
\def\mnras{MNRAS}
\def\Meszaros{M\'esz\'aros } 

\title{The nature of the outflow in gamma-ray bursts}
\author[P. Kumar et al.]{P. Kumar$^1$\thanks{E-mail:
pk@astro.as.utexas.edu}, E. McMahon$^1$,
     A. Panaitescu$^2$, R. Willingale$^3$,
     P. O'Brien$^3$,\newauthor
     D. Burrows$^4$, J. Cummings$^5$,
     N. Gehrels$^5$, S. Holland$^5$,
     S. B. Pandey$^6$,\newauthor
      D. Vanden Berk$^4$, and S. Zane$^6$\\
$^1$Astronomy Department, University of Texas, Austin, TX
78712\\
$^2$Space Science and Applications, MS D466, Los Alamos
National Laboratory, Los Alamos, NM 87545\\
$^3$Department of Physics and Astronomy, University of
Leicester, Leicester, LE 1 7RH, UK\\
$^4$Department of Astronomy and Astrophysics, Pennsylvania
State University, 525 Davey Lab, University Park, PA 16802\\
$^5$NASA Goddard Space Flight Center, Greenbelt, MD 20771\\
$^6$Mullard Space Science Laboratory, University College of
London, Holmbury St Mary, Dorking, Surrey, RH5 6NT, UK
}

\begin{document}
\maketitle

\begin{abstract} { The \emph{Swift} satellite has enabled us to follow
the evolution of gamma-ray burst (GRB) fireballs from the prompt
$\gamma$-ray emission to the afterglow phase. The early x-ray and
optical data obtained by telescopes aboard the \emph{Swift}
satellite show that the source for prompt $\gamma$-ray emission, the
emission that heralds these bursts, is short lived and that its source
is distinct from that of the ensuing, long-lived afterglow. 
Using these data, we determine the distance of the
$\gamma$-ray source from the center of the explosion. We find this
distance to be 10$^{15}-10^{16}$ cm for most bursts and we show that
this is within a factor of ten of the radius of the shock-heated
circumstellar medium (CSM) producing the x-ray photons.  Furthermore,
using the early $\gamma$-ray, x-ray and optical data, we show that the
prompt gamma-ray emission cannot be produced in internal shocks, nor can
it be produced in the external shock; in a more general sense
$\gamma$-ray generation mechanisms based on shock physics have problems
explaining the GRB data for the ten \emph{Swift} bursts analyzed in this
work. A magnetic field dominated outflow model for GRBs has some 
attractive features, although the evidence in its favor is inconclusive.
Finally, the x-ray and optical data allow us to provide an upper limit
on the density of the CSM of about 10 protons per cubic cm at a distance 
of $\sim 5 \times10^{16}$ cm from the center of explosion. }
\end{abstract} 


\section{Introduction}

The x-ray flux of a large fraction of the bursts detected by \emph{Swift} 
exhibits a rapid decline with time, as $\sim t^{-2}$ or faster, for about 10
minutes \citep{tagliaferri05,nousek06,obrien06} after trigger. This is often 
followed by a slowly declining light curve (LC), with the flux falling-off as 
$\sim t^{-1/2}$ for a few hours. The extrapolation of the fast declining x-ray 
LC backward in time matches the LC during the burst, which suggests that the 
early x-ray and late $\gamma$-ray emissions are produced by the same source 
\citep{obrien06}. 

The fastest decline of the LC from a relativistic source moving at
Lorentz factor $\Gamma_0$ and of angular size $\theta_j>\Gamma_0^{-1}$ 
arises when the source switches-off quickly due to, for instance, a rapid 
adiabatic cooling at the end of the ejecta heating episode. In this case, 
the observed flux declines as $t^{-2-\beta}$ \citep{kp00}, where $\beta$ 
is the spectral index of the burst emission, i.e. $f_\nu\propto\nu^{-\beta}$. 
The observed decline rate of the early x-ray LC is often at this theoretical 
limit \citep{obrien06}, therefore the $\gamma$-ray source must have a finite, 
short life and, consequently, must be distinct from the much longer lived 
afterglow source. 

In this paper we determine some properties of the $\gamma$-ray source and
its distance from the center of the explosion using the early time data 
obtained by instruments aboard the Swift satellite.

\section{Gamma-ray source distance}

The early x-ray light curve can be used to determine the distance of the
$\gamma$-ray source ($R_\gamma$) from the central explosion, as suggested
by Lazzati \& Begelman (2006) and Lyutikov (2006). However, instead of
using the unknown GRB jet angle to determine $R_\gamma$, as done in previous
works, we determine the source radius in terms of the forward shock
radius, which has a very weak dependence on the only unknown parameter: 
the density of the circumstellar medium. In order to exploit this method, 
we analyze the $\gamma$-ray, x-ray and optical data within the first 10 minutes 
for ten \emph{Swift} bursts for which we can establish that the steeply falling
off portion of the LC is the large-angle emission. 

Some conditions need to be satisfied by the rapidly falling-off early x-ray 
afterglow LC to be identified with the large-angle emission from the $\gamma$-ray source. 
These conditions are: $(i)$ the temporal decay index ($\alpha$) of the x-ray 
LC during the steep decline phase should be equal to $2+\beta$; $(ii)$ the 
spectral index $\beta$ during early x-ray afterglow should be the same as at 
the end of the gamma-ray burst; $(iii)$ the x-ray afterglow flux extrapolated 
to the end of the prompt $\gamma$-ray emission should be same as the $\gamma$-ray 
flux at the end of the burst extrapolated to the x-ray band. We apply an additional
condition: $t_2/t_1>3$, where $t_1$ \& $t_2$ are the beginning and end of the 
steep x-ray decline phase, to ensure that we have a sufficiently long baseline 
for an accurate determination of $\alpha$ (i.e. this index will not be too
sensitive to the uncertainty in the origin of time).

Ten bursts detected by \emph{Swift} between January 2005 and May 2006
meet these four conditions. Four of these bursts have a single-peaked LC
or are FRED (fast rise, exponential decline) shaped; the remaining six bursts 
contain multiple peaks. The relevant properties for these 10 GRBs are listed
in Table 1.

\begin{table*}
\centering
\begin{minipage}{140mm}
\caption{GRB sample}
\begin{tabular}{llccccccccc}
\hline
GRB & FRED? & $\alpha$ &
$\beta_\gamma$ & $\beta_x$ & $z$ &
$E_{iso,52}$ & $T_{90}$ & $t_2$ & $V$
& $t_{opt}$\\
\hline
050315&yes&4.3$\pm$0.36&1.2$\pm$0.09&1.6$\pm$0.25&
1.95& 8.9& 96& 400& $ >$18.5 &140\\
050713b&yes&3.1$\pm$0.32&0.53$\pm$0.15&0.70$\pm$0.11&
& 23& 120& 720& $>$19.5&190\\
050714b&no&4.8$\pm$1.2&1.7$\pm$0.41&1.7$\pm$0.41&
& 3.0& 70& 550& $>$18.7 &170\\
050814&yes&3.0$\pm$0.17&0.98$\pm$0.19&1.1$\pm$0.08&
5.3& 67& 65& 1300& $>$18.7& 210\\
050819&no&3.0$\pm$0.40&1.6$\pm$0.21&1.2$\pm$0.23&
& 2.2& 36& 900& $>$18.1&130\\
051016a&no&2.93$\pm$1.03&0.95$\pm$0.16&1.2$\pm$0.73&
& 4.5& 22& 530& $>$20.3 &210\\
060108&yes&2.3$\pm$0.31&0.94$\pm$0.11&0.98$\pm$0.25&
2.03& 1.1& 14& 360& $>$19.1 &190\\
060211a&no&3.7$\pm$0.36&0.83$\pm$0.08&0.99$\pm$0.08&
& 7.7& 126& 1000& $>$18 &250\\
060219&no&2.7$\pm$0.75&1.7$\pm$0.28&2.15$\pm$1.06&
& 2.2& 62& 540& $>$18.6&220\\
060223a&no&3.82$\pm$4.84&0.77$\pm$0.08&0.90$\pm$0.23&
4.41& 13& 11& 85& 17.8&190\\
\hline
\end{tabular}
\medskip

$\alpha$ is the decay index of the fast falling early XRT emission,
$\beta_\gamma$ is the BAT spectral index averaged over the duration of the burst, 
$\beta_x$ is XRT spectral index at the beginning of the steep decline phase, 
$z$ is the burst redshift (set to 2.5, the median $z$ for \emph{Swift} bursts,
     for those without a measured $z$), 
$E_{iso,52}$ is the isotropic equivalent energy released in the BAT band (15-150 keV)
  in $10^{52}$ erg, 
$T_{90}$ is the burst duration in seconds ($t_1 \sim T_{90}$ in most cases),
$t_2$ is the time when the steep decline of the x-ray LC ends, measured from $t_0$ as 
    defined in \citet{obrien06},
$V$ is the UVOT magnitude measured at time $t_{opt}$ (in seconds) from the GRB peak.
\end{minipage}
\end{table*}

Consider a $\gamma$-ray source moving at $\Gamma_0$, that turns off at radius 
$R_\gamma$. After the turn-off, the observed x-ray flux comes from regions of 
the $\gamma$-ray source that move at an angle $\theta$ larger than $\Gamma_0^{-1}$ 
with respect to the line of sight \citep{kp00} -- this will be referred to as the 
large-angle emission or {\bf LAE}. The LAE arrives at an observer time 
$t=(1+z)R_\gamma\theta^2/2c$ and has a specific intensity smaller 
than that for $\theta=0$ by factor $(1+\theta^2\Gamma_0^2)^3$. The LAE
starts at $t_1$, the end of the prompt phase, and dominates the LC until 
some time $t_2$ when emission from the forward shock overtakes the rapidly 
decreasing flux from the $\gamma$-ray source. Thus, the source turn-off radius 
is $R_\gamma = 2 ct_1 \Gamma_0^2/(1+z)$.

The 0.3--10 keV fluence of the early rapidly declining x-ray LC, starting from 
the end of the GRB prompt emission to time $t_2$, is greater than $\sim15$\% of 
the GRB fluence for most of the bursts (Table 2). Therefore, the source for the 
steep x-ray LC is not some minor pulse in the explosion but is responsible for 
producing a good fraction of the prompt $\gamma$-ray energy, for both FRED and
non-FRED bursts. For this reason, $t_1$ appearing in the above equation for 
$R_\gamma$ should be roughly equal to the burst duration, $t_\gamma$, otherwise 
the fluence during the LAE would be much less than the observed value.

\begin{table*}
\centering
\begin{minipage}{140mm}
\caption{Calculated quantities} 
\begin{tabular}{lccccccc}
\hline
GRB & LAE fluence$^{(a)}$ & optical flux ratio$^{(b)}$
& $\Gamma_{FS}\left(t_2\right)$ & $R_{FS}\left(t_2\right)^{(c)}$ & $R_\gamma^{(c)}$ \\
\hline
050315  & 0.08 & $2.1\times10^4$ & 84  & 5.8 & 1.4 \\
050713b & 0.08 & $1.8\times10^2$ & 82  & 8.2 & 1.4 \\
050714b & 0.8  & $7.2\times10^4$ & 70  & 4.6 & 0.59 \\ 
050814  & 0.17 & $9.2\times10^2$ & 93  & 11  & 0.53 \\
050819  & 0.66 & $5.0\times10^2$ & 55  & 4.8 & 0.19 \\ 
051016a & 0.47 & $4.9\times10^2$ & 75  & 5.0 & 0.21 \\
060108  & 0.29 & $1.9\times10^1$ & 69  & 3.4 & 0.13 \\
060211a & 0.23 & $2.9\times10^2$ & 63  & 6.7 & 0.86 \\
060219  & 0.4  & $2.8\times10^4$ & 67  & 4.2 & 0.48 \\
060223a & 0.17 &                 & 200 & 3.8 & 0.49 \\
\hline 
\end{tabular}
\medskip 

The first optical data for GRB 060223a was obtained at 187s after the
BAT trigger whereas the steep decline of the x-ray LC ended at 85s
($t_2=85$s). The extrapolation of the x-ray flux at 85s to the optical
band gives a V-mag of 16.3 whereas the observed flux at 187s was 17.8
mag.  For all other bursts UVOT measurements were between $t_1$ and $t_2$. \\
$^{(a)}$ Ratio of fluence from the end of GRB to time $t_2$ in 0.3-10 keV band 
   and the fluence in 15-150 keV band during the burst.
$^{(b)}$ Ratio of the expected to observed optical flux (or upper
limit) at the time of UVOT observation. The expected flux is the
extrapolation of the x-ray flux to the optical band using the XRT
spectral index. 
$^{(c)}$ In units of $10^{16}$ cm. 
\end{minipage}
\end{table*}

The radius ($R_{FS}$) and the LF ($\Gamma_{FS}$) of the shock front in
the CSM are related by $R_{FS}(t_2) \approx 2c t_2 \Gamma_{FS}^2
(t_2)/(1+z)$. Since the energy of the LAE source is a significant
fraction of the total GRB energy, it must have provided a good part of
the kinetic energy deposited in the CSM, thus the LF of the LAE
source, $\Gamma_0$, should be larger than $\Gamma_{FS}$. Given that
$R_{FS}(t_2)/R_\gamma = [\Gamma_{FS}(t_2)/\Gamma_0]^2 (t_2/t_\gamma)$, 
$\Gamma_0 > \Gamma_{FS}(t_2)$ implies that $R_{FS}(t_2)/R_\gamma <
t_2/t_\gamma$. For the ten bursts in our sample, $t_2/t_\gamma$ is
between 5 and 25; the average value of $t_2/t_\gamma$ is 14.0 for the four 
FREDs and 13.5 for the six non-FREDs. If the deceleration time for CSM
shock, $t_d$, is less than $t_2$ (as expected because the x-ray flux is
decreasing monotonically) then the initial LF of the CSM
shock, $\sim\Gamma_0$, is larger than the LF at deceleration by a factor
2, and $R_{FS}(t_d)/R_\gamma$ is smaller than $t_2/t_\gamma$ by a factor
$\sim 4$. Therefore, we conclude that $\gamma$-rays are produced within
a factor $\sim 4$ of the deceleration radius, on the average, for our
sample of bursts.

We now calculate $R_{FS}(t_2)$ and estimate $R_\gamma$. The forward shock
radius at time $t_2$ can be calculated from the dynamics of adiabatic
blast-waves, which yields
$R_{FS}(t_2)=\left[3ct_2 E_{iso}/2 \pi m_p c^2(1+z) n_0 \right]^{1/4}$,
where $E_{iso}$ is the isotropic equivalent of energy in the FS and
$n_0$ is the mean density of the CSM within a sphere of radius
$R_{FS}(t_2)$. The former is obtained from the GRB fluence and the
CSM density (or an upper limit for $n_0$) is calculated from the x-ray and optical
flux at $t_2$.  For the bursts in our sample, we find $n_0\, \lta\, 10\,
{\rm cm}^{-3}$ provided that x-rays are produced via the synchrotron process
(no conditions were imposed on microphysics parameters $\epsilon_e$ and
$\epsilon_B$ in this calculation); the constraint on $n_0$ is weaker if
x-rays are produced via the synchrotron-self-Compton process
\footnote{If the density of the CSM is set by the mass loss from the GRB
progenitor star then this small mean density of $\sim 10$ cm$^{-3}$
along the jet axis, within the radius $R_{FS}(t_2)\sim5\times10^{16}$cm,
means that the mass loss rate divided by the wind speed from the
progenitor star in the polar region, in the last $\sim100$ year of its
life, was smaller than typical Wolf-Rayet stars by at least a factor of
a few 10s.}.  
 From the GRB fluence and assuming $n_0 = 10\, {\rm cm}^{-3}$, we calculate the
forward shock radii $R_{FS}(t_2)$ and LFs, $\Gamma_{FS}(t_2) =
[R_{FS}(t_2)(1+z)/2ct_2]^{1/2}$, given in Table 2. From $R_{FS}(t_2)$, we
calculate the lower bound on the $\gamma$-ray source distance from the
center of explosion and find it to be between 10$^{15}$ and 10$^{16}$cm.
Note that $R_{FS}$ and $\Gamma_{FS}$ have a very weak dependence on
$E_{iso}$ and $n_0$ and therefore any error in $E_{iso}$ or $n_0$ has
small effect on these quantities.

\section{Gamma-ray generation models}

\subsection {Forward-Shock}

Although we find that the burst and early afterglow data are not
incompatible with $R_\gamma \sim R_{FS}(t_2)$, the forward shock (FS)
model for $\gamma$-ray generation can be ruled out because the 
$\gamma$-ray production mechanism is short-lived and because the
FS produces too much optical flux (see below). 
Furthermore, Ramirez-Ruiz \& Granot (2006) have pointed out that the
the relations between the spectral peak, flux and burst duration expected if
$\gamma$-rays are produced in the forward shock are not satisfied
by the GRB prompt emission.

All ten bursts in our sample have deep optical upper limits or
detections a few minutes after the burst -- typically at the beginning
of the steeply declining x-ray LC -- provided by the UV-optical
telescope aboard \emph{Swift}. From the x-ray flux and spectrum 
at the time of the optical observations, we estimate the expected
flux in the optical band and find it to exceed the observed value or
upper limit by two orders of magnitude or more (Table 2). 
A large extinction in the optical can be ruled out because late time 
optical data show it to be less than a factor 2.
Moreover, in those cases with optical detections, the optical spectrum
is consistent with $f_\nu \propto \nu^{-1}$, similar to the spectrum in 
the x-ray band. 
Thus, the deep optical upper limits set by UVOT require that the spectrum 
of the x-ray/$\gamma$-ray source turns over at lower energies and
becomes steeper than $f_\nu \propto \nu^{1/3}$, i.e. that the optical band 
often lies below the synchrotron self-absorption frequency ($\nu_a$) of the 
early x-ray/$\gamma$-ray emission. 
It also implies that the optical flux detected at early times must come from 
a different source.

A straightforward calculation of forward shock emission shows that, if
the x-ray emission at time $t_1$ is produced via the synchrotron process, 
then $\nu_a \ll 2$ eV. This result holds even when we allow for 
an external medium enriched with up to $10^3$ e$^\pm$ pairs per proton.
Therefore, the forward shock model does not satisfy the $\nu_a > 2$ eV 
requirement needed to reconcile the optical and x-ray data at early times. 
If x-rays arise from synchrotron-self-Compton
process then the spectrum below 0.3 keV can be as steep as
$f_\nu \propto \nu$, however, the optical flux associated with the
underlying synchrotron radiation exceeds the observed limit. 

\subsection {Internal Shocks}

We now consider the internal shock model for prompt $\gamma$-ray
generation. According to this model, fluctuations in the LF of the
relativistic outflow lead to collisions between faster and slower ejecta, 
producing internal shocks and $\gamma$-ray radiation. No relationship is 
expected, in general, between where these collisions take place and the 
deceleration radius, whereas we find the average $R_{FS}(t_d)/R_\gamma\lta4$. 
We also found that the average value of $R_{FS}(t_2)/R_\gamma$ is the same 
for bursts with multiple $\gamma$-ray light curve spikes and for FRED bursts. 
This suggests that gamma-rays are produced at a radius that is not set by 
the variability time scale of the central engine, contrary to what is 
expected in the internal shock model. 

The GRB ejecta should consist of baryonic material and/or $e^\pm$ in
order to undergo internal shocks. The interaction of such ejecta with the
CSM launches a reverse shock moving into the ejecta, heating
it and producing synchrotron radiation that peaks in the optical band 
(Panaitescu \& \Meszaros 1998) and  declines with time as $t^{-2}$ 
(Sari \& Piran 1999). It is widely believed that such an emission from 
shocked ejecta was seen for GRBs 990123 and 021211. 

 In Figure 1, we show the early optical light curve for these two bursts 
resulting after subtracting the extrapolation of the late time optical 
emission, which arises in the forward shock. This extrapolation is 
justified because the optical light curves for many Swift bursts 
display a single power-law decline from $\sim 300$ s to hours  
\citep{p06,fan06}. We find that, after subtracting the forward shock
contribution, the early light curves of GRBs 990123 and 021211 decline 
as $f_{opt} \propto t^{-2.5}$. This decline is steeper than expected for
the reverse-shock optical emission and is similar to that of the 
early x-ray LCs.  Therefore, it is likely that the steeply falling early
optical emissions of these bursts are produced via the same mechanism as the
early x-ray, i.e. the LAE from the $\gamma$-ray source (Panaitescu et 
al. 2006). This interpretation is also supported by the observations
that for both these bursts the prompt emission spectrum below the peak 
is $f_\nu \propto \nu^{1/3}$. 

Furthermore, a good
fraction of Swift bursts have been followed in the optical starting at a
few minutes after the burst and most of these have either weak optical
flux or very stringent upper limit on the flux (Roming et al.  2005).
Therefore, we lack evidence for the expected reverse-shock emission
from a baryonic/leptonic ejecta. There are various possibilities to
account for a dim reverse-shock emission including the obvious one that
there is no reverse shock because the baryonic/leptonic component in GRB
outflows is small and the bulk of the explosion energy is carried
outward by magnetic fields. 

\begin{figure*}
\includegraphics[width=13cm]{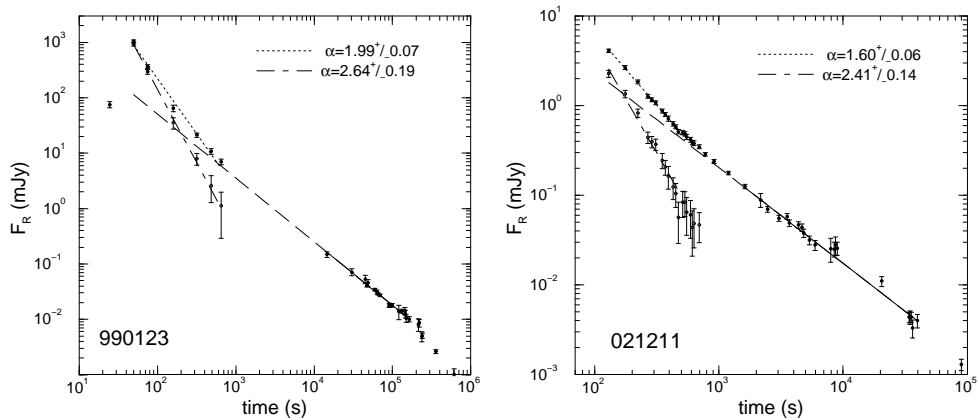}
\caption
{{\bf Left panel}: Power-law fits to the early and late optical
afterglow of GRB 990123. Dotted line shows a power-law fit to the ROTSE 
data at $50-10^3$ s after trigger, solid line is the power-law fit 
($\alpha= 1.15 \pm 0.07$) to the forward-shock
emission at $10^4-10^5$ s, which is back-extrapolated (dashed line) 
to the epoch of the ROTSE measurements. Dot-dashed line shows the
fit to the ROTSE emission with the forward-shock subtracted -- 
the residual flux declines as $t^{-2.64 \pm 0.19}$. 
 {\bf Right panel}: Power-law fits to the early and
late optical afterglow of GRB 021211. Dotted line shows the fit to the
KAIT data at 100--500 s, solid line is the fit ($\alpha = 1.07 \pm 0.04$)
to the forward-shock emission at $10^3-4\times 10^4$ s,
which is back-extrapolated (dashed line) to the epoch of the early KAIT
measurements. Dot-dashed line shows the fit to the KAIT emission with
the forward-shock subtracted  -- the residual flux decays as
$t^{-2.41 \pm 0.14}$. }
\label{fig:noRS}
\end{figure*}

\subsection {Modeling GRB prompt emission}

We can obtain further insights regarding $\gamma$-ray sources by
modeling the average properties of the prompt emission in our set of
GRBs.  The basic procedure is to calculate the synchrotron and IC
radiations for a relativistic, shock heated medium and compare this to
the average burst spectrum and variability timescale.  This synchrotron
and IC radiation is completely described by five parameters: $B$,
$\tau_e$, $\Gamma_0$, $N_e$, and $\gamma_i$, which are respectively,
magnetic field strength, optical depth of the source to Thompson
scattering, the LF of the source, the total number of shocked electrons,
and the lowest LF of electrons in the source comoving frame just behind
the shock front; the electron distribution just behind the shock front
is a power-law function of index $p$ which is constrained by the observed
high energy spectra.  The distribution in the source as a whole has a
more complicated shape due to radiative losses which we calculate using
the five parameters.  We determine which part of the 5D parameter space
produces radiation matching the observed low energy spectral index, peak
energy, flux at the peak, and average pulse duration of the GRBs in our
sample.  The solutions we find apply to any relativistic-shock
heated medium -- internal or external shocks.

We first attempt to describe the prompt emission of these 10 bursts with
synchrotron radiation. The low energy (20-150 keV) spectral index for 6
of the 10 bursts is $0.5<\beta_\gamma< 1$, and therefore the synchrotron
cooling frequency ($\nu_c$) should be larger than about 150 keV and the
injection frequency $\nu_i$ below 20 keV.  This constraint along with
peak flux of 0.2 mJy and pulse duration of 10s produces a 5D solution
space with $\Gamma_0 > 600$ and $R_\gamma = \left(N_e \sigma_T/4 \pi
\tau_e\right)^{1/2} \gta 10^{17}$ cm (fig. 2a).  This is in
contradiction to what we found using the steep x-ray light curve decay
-- $R_\gamma \lta 10^{16} $cm and bulk LF of $< 100$ (table 2).  This
discrepancy suggests that synchrotron radiation from a relativistically
shock heated medium (internal or external shocks) cannot describe the
prompt $\gamma$-ray emission properties of the GRBs in our sample.  For
the remaining four GRBs, $1.2<\beta_\gamma<1.8$ and both $\nu_i$ and
$\nu_c$ should be below 20 keV.  The synchrotron solutions for this case
for the most part are very similar to the previous synchrotron case.
There are a few intriguing solutions consistent with the $R_\gamma$ and
$\Gamma_0$ found in the LAE calculation, but the prompt optical flux is
very bright, and can also be ruled out.  Therefore, we rule out
synchrotron emission in shock heated medium as the mechanism for GRB
prompt emission\footnote{Three assumptions were made in these
calculations: electron pitch angle distribution is uniform; electrons
are not continuously energized as they move downstream from the shock
front; and $B$ does not vary by a large factor across the source.}.

\begin{figure*}
\includegraphics[width=13cm]{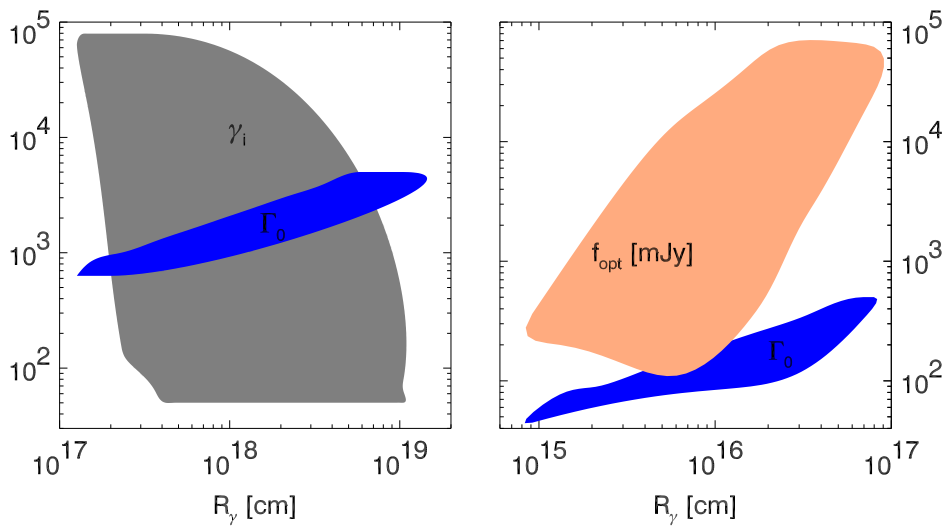}
\caption
{{\bf Left panel}: the allowed range of value for $R_\gamma$,
$\Gamma_0$ (the LF of the $\gamma$-ray source -- blue band) and
$\gamma_i$, the minimum LF of shocked electrons close to the shock
front, for the case when the prompt GRB emission is produced via the
synchrotron process. These results were obtained for a GRB pulse
duration of 10 s, the flux at 100 keV of 0.2 mJy, cooling frequency
($\nu_c$) greater than 150 keV and the synchrotron frequency $\nu_m$
corresponding to $\gamma_i$ less than 20 keV, so that the
spectrum in the BAT band corresponds to $f_\nu \propto\nu^{-(p-1)/2}$.
For a GRB pulse duration of 1 s the minimum $R_\gamma$ decreases by a
factor of $\sim 4$ and the minimum $\Gamma_0$ increases by a factor of
$\sim 2$. The allowed parameter space for synchrotron solution is found
to be not very sensitive to the peak flux, $\nu_c$ and $\nu_m$. The
allowed range for $R_\gamma$ \& $\Gamma_0$ is very similar 
for the $\gamma$-ray fluxes measured for the 10 bursts in our sample,
including the  $\nu_m<\nu_c<20$ keV (i.e. $f_\nu\propto\nu^{-p/2}$) case.
For $\nu_c<\nu_m<20$ keV, there are solutions consistent with the parameters
shown in Table 2, but they lead to a too bright optical flux. 
The large range allowed for $\gamma_i$ encompasses internal and external 
shock `solutions'. 
{\bf Right panel}: 
The allowed range of values for $R_\gamma$ and $\Gamma_0$ in the case
when the burst emission is synchrotron self-Compton case and for the same 
burst parameters as for the left panel. Also shown is the optical
flux (in mJy) for the SSC solutions. 1mJy corresponds to an R-magnitude of
16.2; the upper limits on the optical flux for most GRBs in our sample
is $\lta 0.1$ mJy.  }
\label{fig:paras}
\end{figure*}

Is it possible that the $\gamma$-rays were produced via
synchrotron-self-Compton (SSC) process in a relativistic shock?  We
perform the 5D parameter space search for SSC radiation for both of the
$\beta_\gamma$ cases described above and find that (for either
$\beta_\gamma$) the source radius $R_\gamma$ and $\Gamma_0$ for the
allowed 5D parameter space are consistent with the values we obtained
for our sample in table 2 (see fig. 2b).  The problem, however, is that
the prompt optical flux with SSC is many orders of magnitude larger than
the observational upper limits (fig.  2b).  It is very unlikely that
this large flux has gone undetected because of dust extinction or bursts
going off at very high redshifts \citep{r06}.  Therefore, we
conclude that GRB prompt emission is not due to the SSC process in
relativistic shocks either.  This means that synchrotron or SSC from any
shock model has problems describing the $\gamma$-ray emission in any of
the bursts in our sample -- and that internal \& external shocks can be 
ruled out as possible $\gamma$-ray emission mechanisms.

We have described a few problems with the external and internal
shock models and, more generally, for any model based on shock physics.
These together with the lack of evidence for baryonic outflow -- no firm
detection of reverse-shock emission in GRBs -- suggests that GRB prompt
emission is produced by some very different process. It either involves
a very different kind of shock physics than we see during GRB
afterglows, which seems unlikely, or $\gamma$-ray generation does not
involve shocks, such as, for instance, would be the case when magnetic
field transports the energy in GRB outflows and its dissipation produces
the radiation we see cf. \citet{usov92,usov94}, \citet{thompson94},
\citet{katz97}, \Meszaros \& Rees (1997), Wheeler et al. (2000 \& 2002), 
Vlahakis \& Konigl (2001), Spruit et al.  (2001), Lyutikov \& 
Blandford (2003).  The Poynting model has some attractive features such as
high radiative efficiency, no reverse shock, large radius for
$\gamma$-ray source (Lyutikov \& Blandford, 2003), and low baryon
loading comes for free.  The Poynting outflow, however, might have
difficulty explaining the  observed variability of GRB prompt light curve
(personal communication, Piran).

\section{Summary}

The early x-ray data show that the gamma-ray source is short lived and
turns off at a distance of a $\sim 5\times10^{15}$ cm from the central
explosion -- which is found to be within a factor of $\sim10$ of the
forward shock radius at early times for all ten bursts in our sample.
We have presented arguments that the prompt
$\gamma$-ray emission is unlikely to be produced in the external or
internal shocks or any mechanism based on shock heating of electrons.
In their electromagnetic model, Lyutikov \& Blandford (2003) find that 
$\gamma$-rays are generated at a distance of $\sim3\times 10^{16}$ cm 
from the central explosion, which is comparable to the value that we find. 
This could just be a coincidence but, considering the problems with shock 
based models, the lack of reverse-shock optical detection, and very high
efficiency for $\gamma$-ray generation, we find the Poynting outflow
model for GRBs to be an attractive possibility. 

\section*{Acknowledgments}

This work is supported in part by grants from NASA
and NSF (AST-0406878) to PK.  We thank Tsvi Piran for helpful comments
and Craig Wheeler for useful discussions on this work.

\end{document}